\documentclass[twocolumn,trackchanges]{aastex7}

\newcommand\Msun{M$_\odot$\,}
\newcommand\Rsun{R$_\odot$\,}
\newcommand\Lsun{L$_\odot$\,}
\graphicspath{{./}{figures/}}

\begin{document}


\title{Fading into darkness: A weak mass ejection and low-efficiency fallback accompanying black hole formation in M31-2014-DS1}

\author[orcid=0000-0002-8989-0542, sname='Kishalay De']{Kishalay De} 
\affiliation{Department of Astronomy and Columbia Astrophysics Laboratory, Columbia University, New York, NY, USA}
\affiliation{Center for Computational Astrophysics, Flatiron Institute, New York, NY, USA}
\email{kd3038@columbia.edu}

\author[0000-0002-1417-8024]{Morgan MacLeod} 
\affiliation{Center for Astrophysics | Harvard \& Smithsonian, Cambridge, MA, USA}
\email{morgan.macleod@cfa.harvard.edu}

\author[]{Jacob E. Jencson}
\affiliation{IPAC, California Institute of Technology, Pasadena, CA, USA}
\email{}

\author[]{Ryan M. Lau}
\affiliation{IPAC, California Institute of Technology, Pasadena, CA, USA}
\email{}

\author[]{Andrea Antoni}
\affiliation{Center for Computational Astrophysics, Flatiron Institute, New York, NY, USA}
\email{}

\author[orcid=0000-0002-5296-6232,gname='Maria Jose']{Mar\'ia Jos\'e Colmenares}
\affiliation{Department of Astronomy, University of Michigan, Ann Arbor, MI, USA}
\email{}

\author[0000-0001-6947-6072]{Jane Huang}
\affiliation{Department of Astronomy and Columbia Astrophysics Laboratory, Columbia University, New York, NY, USA}
\email{}

\author[]{Megan Masterson}
\affiliation{Kavli Institute for Astrophysics and Space Research, Massachusetts Institute of Technology, Cambridge, MA, USA}
\email{}

\author[0000-0003-2758-159X]{Viraj R. Karambelkar}
\altaffiliation{NASA Hubble Fellow}
\affiliation{Department of Astronomy and Columbia Astrophysics Laboratory, Columbia University, New York, NY, USA}
\email{}

\author[]{Mansi M. Kasliwal}
\affiliation{Cahill Center for Astrophysics, California Institute of Technology, Pasadena, CA, USA}
\email{}

\author[]{Abraham Loeb}
\affiliation{Center for Astrophysics | Harvard \& Smithsonian, Cambridge, MA, USA}
\affiliation{Black Hole Initiative, Harvard University, Cambridge, MA, USA}
\email{}

\author[]{Christos Panagiotou}
\affiliation{Kavli Institute for Astrophysics and Space Research, Massachusetts Institute of Technology, Cambridge, MA, USA}
\email{}

\author[]{Eliot Quataert}
\affiliation{Department of Astrophysical Sciences, Princeton University, Princeton, NJ, USA}
\email{}

\correspondingauthor{Kishalay De}
\email{kd3038@columbia.edu}

\begin{abstract}
Stellar-mass black holes (BHs) can form from the near-complete collapse of massive stars, causing them to abruptly disappear. The star M31-2014-DS1 in the Andromeda galaxy was reported to exhibit such a disappearance between 2014 and 2022, with properties consistent with the failed explosion of a $\approx 12 - 13$~M$_\odot$ yellow supergiant leading to the formation of a $\approx 5$\,\Msun BH. We present mid-infrared (MIR) observations of the remnant obtained with the \textit{James Webb Space Telescope} (JWST) and X-ray observations from the \textit{Chandra X-ray Observatory} in 2024. The JWST MIRI/NIRSpec data reveal an extremely red source, showing strong blueshifted absorption from molecular gas (CO, CO$_2$, H$_2$O, SO$_2$) and deep silicate dust features. Modeling the dust continuum confirms continued bolometric fading of the central source to $\log(L/L_\odot)\approx3.88$ ($\approx7$--8\% of the progenitor luminosity), surrounded by a dust shell spanning $\approx40-200$~au. Modeling of the molecular gas indicates $\sim 0.1$~M$_\odot$ of gas expanding at $\approx 100$\,km\,s$^{-1}$ near the inner edge of the dust shell. No X-ray source is detected down to a luminosity limit of $L_X\lesssim1.5\times10^{35}$~erg~s$^{-1}$. We show that the panchromatic observations are explained by (i) a low-energy ($\approx10^{46}$~erg) ejection of the outer H-rich progenitor envelope and (ii) a fading central BH powered by inefficient ($\sim0.1$\% in mass) accretion of loosely bound fallback material. The analysis robustly establishes the bolometric fading of M31-2014-DS1 following its $> 30$\,yr archival record of constant optical/infrared brightness, and provides the first cohesive insights into BH formation via low-energy explosions and long-term fallback.
\end{abstract}



\section{Introduction}

Despite decades of theoretical work and widespread observational evidence, the formation channels of stellar-mass black holes (BHs) remain poorly constrained by direct observations. BHs are understood to represent the end states of massive stellar evolution, with theory predicting multiple pathways to their formation \citep{Burrows2025, Janka2025}. These include both successful terminal explosions -- in which the progenitor star ejects most of its core and outer envelope in a core-collapse supernova (CCSN; \citealt{Janka2007, Smartt2009}) -- and failed explosions, where the star undergoes near-complete implosion with little accompanying mass ejection \citep{Kochanek2008, Lovegrove2013}. While CCSNe are now routinely observed in wide-field transient surveys (e.g. \citealt{Perley2020, Pessi2025}), identifying failed SNe and their weak mass-loss signatures has proven far more challenging, leaving this channel largely unconstrained \citep{Kochanek2008, Byrne2022, Basinger2021}. Due to their expected low explosion energies ($\sim 10^{45} - 10^{49}$\,erg), the most dramatic observational signature of failed SNe has been through the identification of luminous supergiants in nearby galaxies that disappear without an observed CCSN \citep{Adams2017, De2026}.

The first promising candidate was presented by \citet{Adams2017} who reported such an identification of NGC\,6946-BH1 in the galaxy NGC\,6946, which exhibited a brief ($\lesssim 300$\,d), low luminosity $\sim 10^6$\,\Lsun outburst prior to its abrupt optical disappearance. Combining more than decade of optical photometry from the Large Binocular Telescope \citep{Gerke2015, Basinger2021}, its properties were shown to be consistent with the failed explosion of a massive ($\gtrsim 20$\,\Msun), luminous ($\log (L/L_\odot) \approx 5.3$) supergiant where the remnant was subsequently enshrouded in dust formed from the low energy mass ejection \citep{Kochanek2014, Kochanek2024}. More recently, \citet{Beasor2023} and \citet{Kochanek2023} presented follow-up mid-infrared (MIR) observations of the remnant obtained with the {\it James Webb Space Telescope} (JWST), which reveal a lingering MIR source at $\approx 20$\% of the progenitor luminosity. Although its observational appearance was suggested to be similar to the dusty remnants of stellar mergers \citep{Beasor2023}, the unambiguously lower bolometric luminosity was argued to support a terminal disappearance of the progenitor \citep{Kochanek2023, Kochanek2024}.

While the conclusions regarding NGC\,6946-BH1 were limited by the sparse temporal coverage of this (relatively) distant star, (\citealt{De2026}; hereafter D26) presented a new example of this phenomenon in the fortuitously nearby Andromeda galaxy. As in the case of NGC\,6946-BH1, the source M31-2014-DS1 was identified as a luminous $\approx 12-13$\,\Msun yellow supergiant (YSG) that disappeared in optical light between 2014 and 2018, with exquisite MIR coverage of the fading from the NEOWISE mission. By combining a panchromatic archival dataset, D26 demonstrated that the fading amplitude and timescale was well explained by the hydrogen-depleted progenitor envelope -- that was partially initially expelled and subsequently accreted during the collapse. They showed that this ejection and partial fallback model also explains the observed properties of NGC\,6946-BH1 -- with its optical outburst and fading timescale indicative of a $\approx 17.5$\,\Msun YSG progenitor star with a $\approx 0.6$\,\Msun H-rich envelope, as previously also suggested from archival photometry \citep{Humphreys2019}. No neutrinos have been detected from a targeted search towards M31-2014-DS1 \citep{Nakanishi2025, Suwa2025}.

Unlike prior observations of NGC\,6946-BH1 which were limited to broad-band photometry, the remarkable proximity of M31-2014-DS1 offers the ability to carry out exquisite follow-up of the MIR remnant as well as sensitive X-ray observations to test the presence of an accreting BH. In this paper, we present MIR spectroscopy and imaging of the remnant of M31-2014-DS1 in 2024 obtained with JWST, together with deep follow-up observations with the {\it Chandra} X-ray Observatory (CXO). The structure of this paper is as follows. In Section \ref{sec:observations} we describe the JWST and \emph{Chandra} observations and data reduction. Section \ref{sec:analysis} presents an empirical comparison of the data to other dusty transients, together with quantitative modeling of the dust continuum and molecular absorption features. In Section \ref{sec:modeling} we interpret the observational results in the context of mass ejection and fallback accretion models for failed SNe. We summarize our conclusions, discuss broader implications and the future evolution in Section \ref{sec:summary}. For the entire paper, we adopt a distance of $770$\,kpc to M31 \citep{Savino2022}. 

\section{Observations}
\label{sec:observations}

\subsection{JWST Observations}

We obtained observations of M31-2014-DS1 with JWST as part of our Director's Discretionary Time Program (DD 6809; PI: De). Observations were carried out between UT 2024-11-28 03:13 and 2024-11-28 10:55 using the Mid-infrared Instrument (MIRI; \citealt{Wright2023}) and the Near-infrared Spectrograph (NIRSpec; \citealt{Boker2023}). The MIRI observations consisted of a $R\sim 100$ spectrum covering $5-12\,\mu$m obtained with the Low Resolution Spectrometer (LRS) and broadband imaging observations in the F1280W, F1500W, F1800W, F2100W and F2550W filters. The NIRSpec observations consisted of a medium resolution $R\sim 3000$ spectrum obtained in the G140H, G235H and G395H gratings. Both the MIRI LRS and NIRSpec observations included an offset target acquisition due to the expected faintness of the source at $\sim 3-5\,\mu$m  and the very crowded field in the near-infrared bands (D26; their Figure 1).

\begin{figure*}[!ht]
    \centering
    \includegraphics[width=0.49\linewidth]{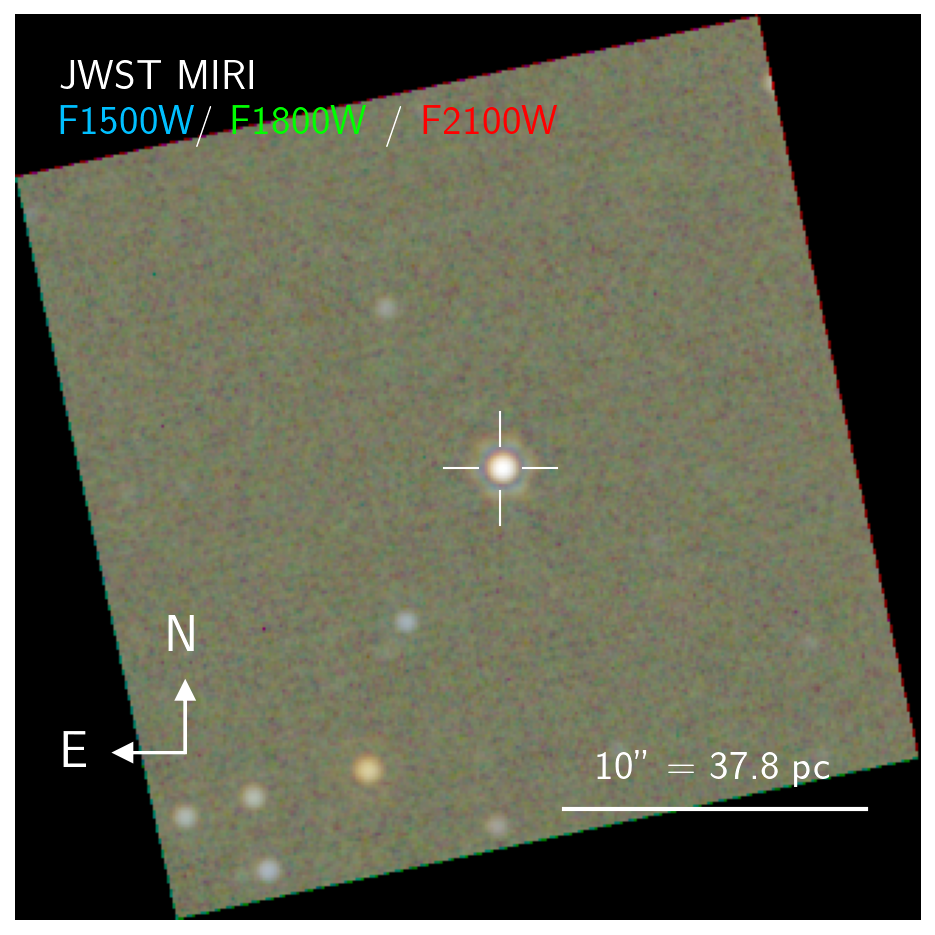}
    \includegraphics[width=0.49\linewidth]{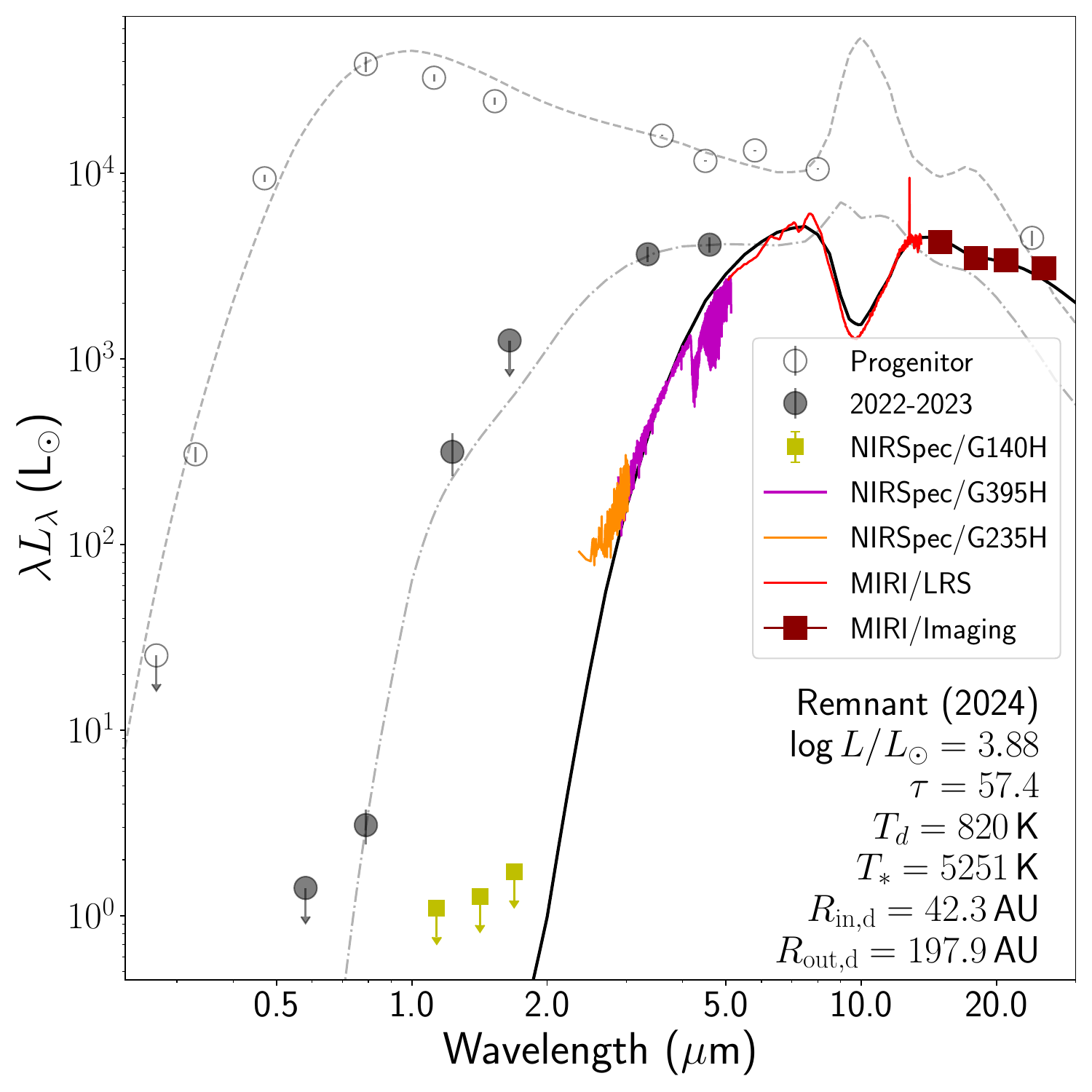}
    \caption{(Left) JWST/MIRI RGB composite image of M31-2014-DS1 (color channels shown in the label). The spatial scale and orientation of the image are shown, and the source position is marked with a white crosshair. (Right) The evolution of the SED of M31-2014-DS1. The empty circles show the progenitor SED measured from 2005-2012 data, with the dashed line showing the best-fit \texttt{DUSTY} model. The filled circles show the SED of the source from 2022-2023, along with its \texttt{DUSTY} model as dot-dashed lines. The JWST data from NIRSpec, MIRI LRS and MIRI imaging from December 2024 are shown as colored lines (when detected at $>10\sigma$ significance; see legend), along with its best-fit \texttt{DUSTY} model as black lines. The source was not detected in NIRSpec/G140H observations; we show $10\sigma$ upper limits binned to $0.3\,\mu$m intervals. Some best-fit parameters derived from dust continuum modeling of the JWST data are shown (see also Table \ref{tab:params}).}
    \label{fig:spec}
\end{figure*}

Calibrated data were downloaded from the MAST portal. The MIRI and NIRSpec spectroscopic data were reduced using version 1.20.2 of the JWST science calibration pipeline and calibrated using version 13.0.6 of the CRDS with the 1464 CRDS context. Photometry from the MIRI imaging observations were extracted by performing aperture photometry using an aperture that encloses 70\% of the total flux and corresponding aperture corrections. The source is clearly detected in all the MIRI observations, and a multi-color RGB composite of the imaging data are shown in Figure \ref{fig:spec}. The source is clearly detected in G395H and G235H observations; however, the extremely red source becomes exceedingly faint at shorter wavelengths in G235H and is undetected in the G140H data. In addition to using the (partial) G235H and G395H spectra, we use the non-detection in the G140H observations to place upper limits on the source spectral energy distribution (SED). The combined SED of the source is shown in Figure \ref{fig:spec}.

\subsection{Chandra X-ray Observations}

We obtained follow-up X-ray observations of M31-2014-DS1 using the {\it Chandra} Advanced CCD Imaging Spectrometer-S \citep{Garmire2003} as part of our Director's Discretionary program (Seq 503585; PI: De). Observations were executed on UT 2024-11-06, 2024-11-08 and 2024-11-09 for a total exposure time of $50$\,ks.  We reduced the data using \textsc{Ciao} \citep[version 4.16;][]{Fruscione2006} and CALDBv4.11.5 \citep{CXCCALDB}, finding no source at the Gaia position. Using the \texttt{aplimits} tool \cite{Fruscione2006} and following published procedures\footnote{\url{https://cxc.cfa.harvard.edu/ciao/threads/upperlimit/}}, we estimate a 3$\sigma$ upper limit on the X-ray flux $F_X \lesssim 2.1 \times 10^{-15}$~erg~s$^{-1}$~cm$^{-2}$ in the 0.5-7 keV band. This limit was calculated using a circular source region with a radius of 5\arcsec\ and by assuming a power-law spectrum (number of photons $N \propto E^{-\Gamma}$, where $E$ is the photon energy and $\Gamma = 2$ is the assumed photon index) with Galactic absorption\cite{HI4PICollaboration2016} corresponding to a hydrogen column density $N_H = 2.26 \times 10^{21}$~cm$^{-2}$. This translates to an X-ray luminosity $L_X \lesssim 1.5 \times 10^{35}$~erg~s$^{-1}$ in the 0.5-7 keV band.

\section{Analysis}
\label{sec:analysis}
\subsection{Comparison to remnants of dusty transients}

\begin{figure*}[!ht]
    \centering
    \includegraphics[width=0.49\linewidth]{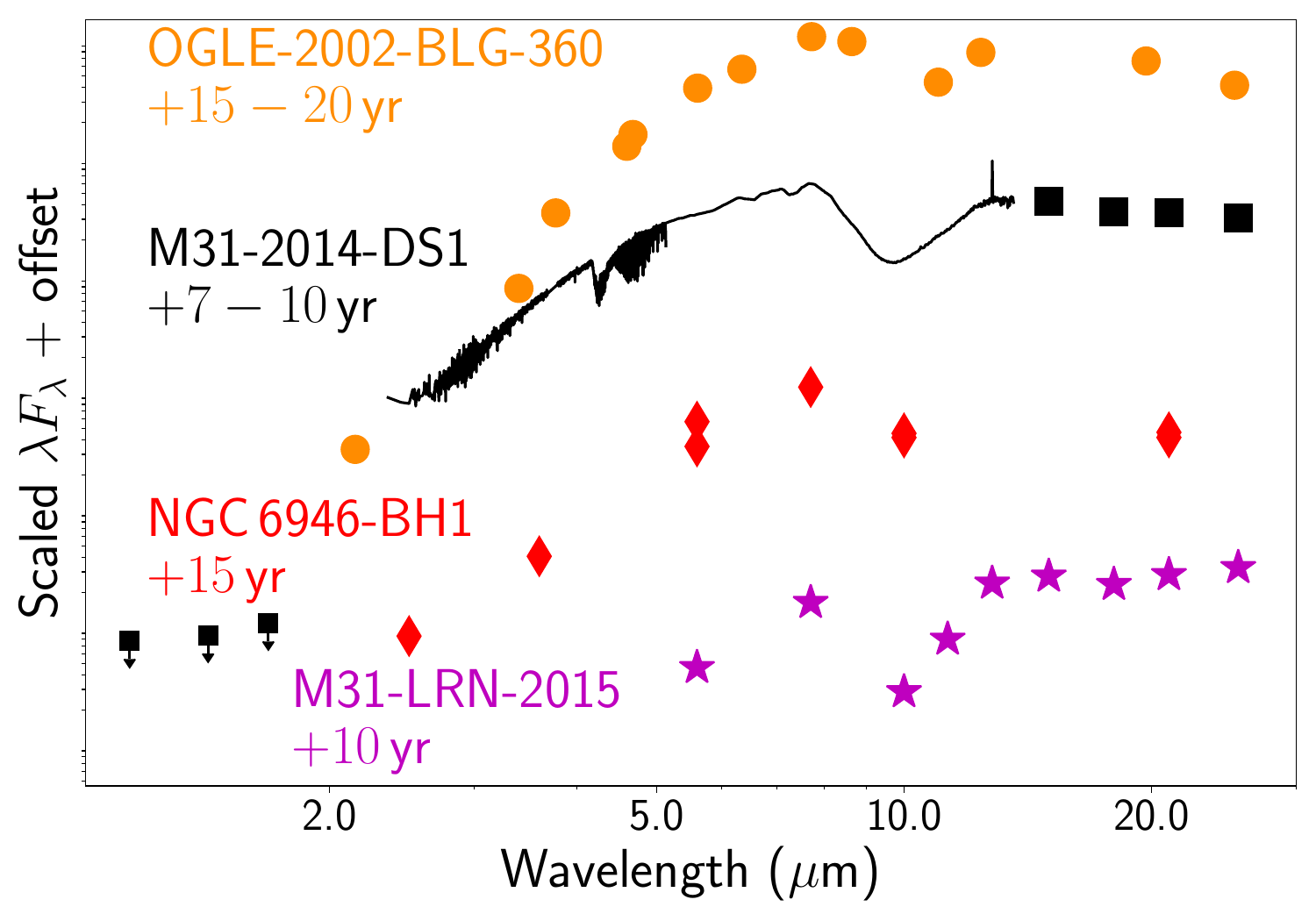}
    \includegraphics[width=0.49\linewidth]{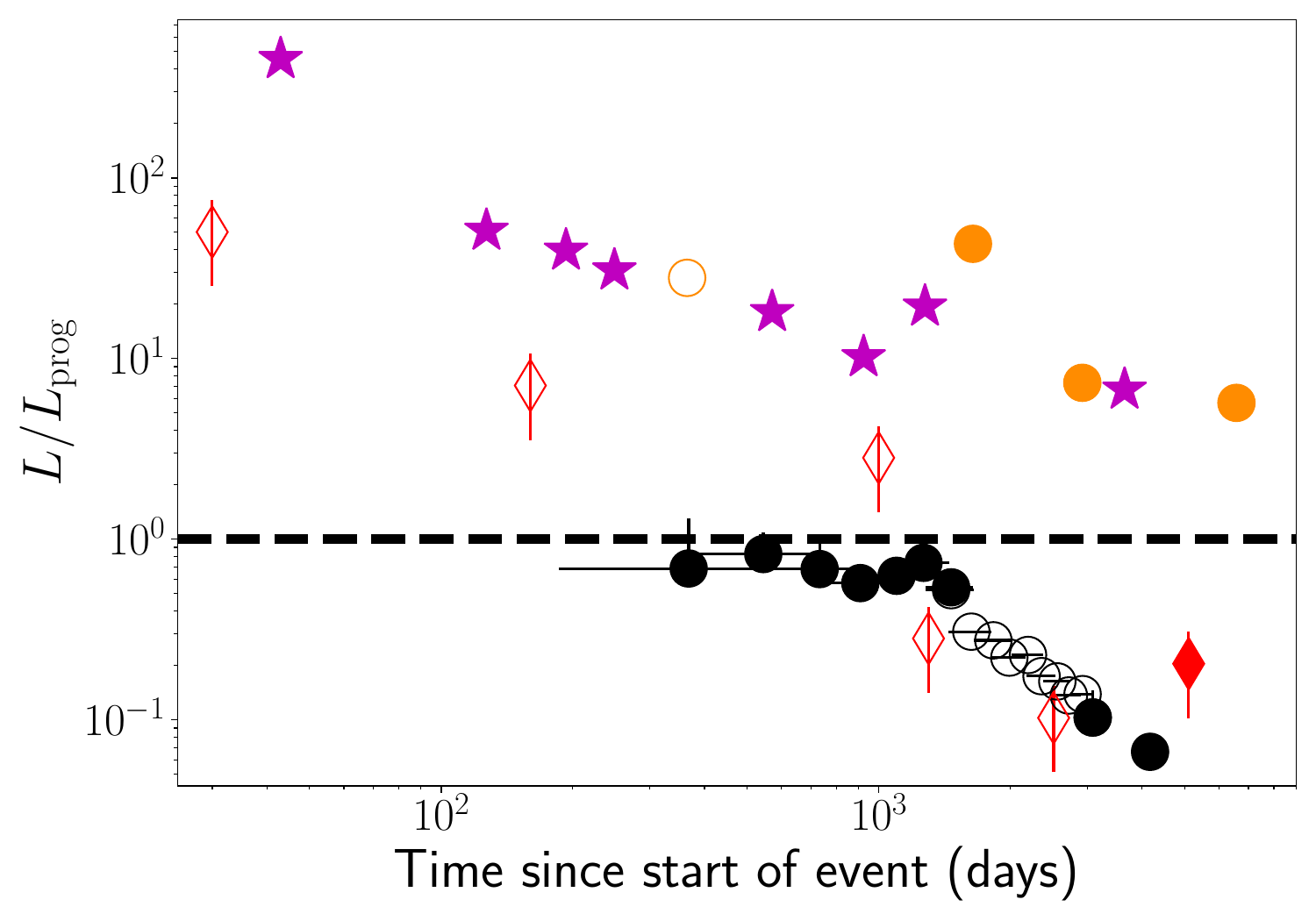}
    \caption{Comparison of M31-2014-DS1 to other dusty transients. ({\it Left}) Comparison of the MIR JWST spectrum of M31-2014-DS1 to other dusty transients (indicated in the same color as the text along with the phase of observation) -- an optically discovered luminous red nova (M31-LRN-2015; \citealt{Karambelkar2025, Blagorodnova2020}), a dusty Galactic stellar merger from an evolved primary star (OGLE-2002-BLG-360; \citealt{Steinmetz2025, Tylenda2013}) and the failed SN candidate NGC\,6946-BH1 \citep{Beasor2023, Kochanek2023, Adams2017}. The flux scales have been arbitrarily shifted for visualization. ({\it Right}) Comparison of the bolometric light curves of the sources relative to the estimated progenitor luminosity $L_{\rm prog}$ (the dashed line indicates a ratio of unity), with the same labeling scheme. In the case of M31-2014-DS1, we show the calculated pseudo-bolometric luminosity by integrating the SED over the observed wavelength range ($1-30\,\mu$m; Figure \ref{fig:spec}). Solid markers indicate epochs that included complete optical to mid-IR SED modeling, while hollow markers indicate epochs without complete spectral coverage.}
    \label{fig:comp}
\end{figure*}

The MIR SED of the M31-2014-DS1 exhibits some distinctive features -- i) an extremely red slope, with the spectrum rising by $\gtrsim 3000\times$ in luminosity from $2 - 8\,\mu$m, ii) a forest of strong, narrow absorption lines between $4-5\,\mu$m in the NIRSpec data, iii) a broad, deep absorption trough between $8-13\,\mu$m in the MIRI LRS data, which we identify with silicate absorption and iv) a fading SED at longer wavelengths inferred from the MIRI broadband imaging observations. The silicate absorption confirms the assumption of D26 of O-rich circumstellar chemistry surrounding the star (which was previously not directly constrained). These characteristics are similar in many ways to those of dusty remnants of other types of low-energy transients and generally to dusty gas in O-rich circumstellar environments. Figure \ref{fig:comp} compares the observed SED of M31-2014-DS1 to previously published observations of the remnant of a stellar merger in M31 (M31-LRN-2015 \citealt{Karambelkar2025}), the dustiest known Galactic stellar merger (OGLE-2002-BLG-360; \citealt{Steinmetz2025}), and the remnant of the failed SN candidate NGC\,6946-BH1 (\citealt{Beasor2023, Kochanek2023}).

The MIR spectrum of M31-2014-DS1 is remarkably similar to these events in their general red color as well deep silicate absorption near $\approx 10\,\mu$m. The similarity is particularly relevant for the case of NGC\,6946-BH1, which was suggested by D26 to be a failed SN from a marginally more massive progenitor with a larger H-envelope. Since MIR spectroscopy of its remnant are not available, \citet{Beasor2023} suggested that the photometric excess at $\approx 8\,\mu$m in NGC\,6946-BH1 could be explained by polycyclic aromatic hydrocarbon (PAH) features, while \citet{Kochanek2023} argued that it was associated with silicate absorption from dust in O-rich gas. A comparison of M31-2014-DS1 and NGC\,6946-BH1 suggests that the MIR SED of NGC\,6946-BH1 is consistent with deep silicate absorption expected from an O-rich dust obscured source and generally the circumstellar environments of massive stars \citep{Verhoelst2009}. We attribute the similarity of the remnants of both M31-2014-DS1 and NGC\,6946-BH1 to those of the stellar mergers as the characteristic appearance of dust obscured stars in O-rich environments.

Figure~\ref{fig:comp} also compares their bolometric luminosity evolution relative to progenitor values. For M31-2014-DS1, we adopt the published bolometric light curve through 2022 from D26 and extend it with the JWST epoch by performing a trapezoidal integration of the observed spectrum. This pseudo-bolometric luminosity is computed by integrating the observed SED (Figure \ref{fig:spec}) between $1-30\,\mu$m, and performing a linear extrapolation to the edges of the wavelength range where the source falls below the detection threshold. In contrast to the failed SN candidates, the stellar mergers unambiguously leave remnants that remain over-luminous with respect to their progenitors for decades after eruption \citep{Tylenda2013, Tylenda2016, Woodward2021}. The failed SN candidates instead exhibit substantially lower relative luminosities, supporting their association with terminal disappearance. While \citet{Kashi2017} have qualitatively suggested that the apparent remnant luminosity can be suppressed due to obscuration by a dusty disk, \citet{Kochanek2024} have quantitatively shown that this suppression can be at most  a factor of $\approx 2\times$. Although the MIR coverage of NGC,6946-BH1 is sparse to construct a well-sampled bolometric light curve, the densely sampled NEOWISE data for M31-2014-DS1, combined with the new JWST observations, demonstrate continued fading to $\lesssim 10$\% of the progenitor luminosity.

There have been $\approx 3$ low-mass stellar mergers reported from optical searches in M31 in the last 40 years \citep{Rich1989, Lipunov2017, Pastorello2021}. \citet{Karambelkar2023} estimate that higher mass mergers ($\gtrsim 13\,$\Msun primary stars, relevant to the progenitor of M31-2014-DS1) are $10-100\times$ rarer than lower mass objects. Therefore, the likelihood of detecting a detecting a failed SN (estimated $\approx 1-20$\% by D26) is no less than that for a high mass merger is M31 over the NEOWISE time baseline ($\approx 1-10$\% over $15$\,yrs). This can be understood by noting that the fraction of massive stars estimated to undergo mergers ($\approx 25$\%; \citealt{Sana2012}) is very similar to the estimated fraction that undergo a failed SN \citep{Neustadt2021}.

\citet{Beasor2023} note similarities between the IR-luminous remnant of NGC\,6946-BH1 and that of the post-eruption fading from the Great Eruption of Eta Carinae \citep{Smith2011} -- suggesting a possible analog if the progenitor of NGC\,6946-BH1 was in an elevated luminosity state prior (e.g. due to ongoing binary interaction) to outburst. However, the exquisite archival data on M31-2014-DS1 allows us to confirm that i) the progenitor of M31-2014-DS1 has remained at nearly constant optical/infrared brightness since its earliest published reports from observations in 1980 - 1990 ($i \approx 17.5$\,mag for $>30$\,yrs before the disappearance; \citealt{Monet2003, Magnier1992, Skrutskie2006}), and ii) unlike the Eta Carinae remnant that has maintained a near-constant bolometric luminosity at the progenitor level for decades post-eruption \citep{Mehner2019}, M31-2014-DS1 continues to fade in luminosity (Figure \ref{fig:comp}).

\subsection{Dust continuum modeling}

M31-2014-DS1 exhibits a clear peak of emission between $5-20\,\mu$m as expected from warm dust surrounding the remnant. To quantify the properties of the associated dust, we model the observed SED using the radiative transfer code \texttt{DUSTY} \citep{Ivezic1997}. We assume a spherically symmetric dust shell surrounding a central source described by a blackbody spectrum, together with warm astronomical silicate dust grains (given the observed deep absorption at $\approx 8-12\,\mu$m) with a single characteristic grain size and a $\rho \propto r^{-2}$ density profile. The resulting free parameters of the dust model include the bolometric flux ($F$; or equivalently luminosity $L$) and effective temperature ($T_*$) of the central source, the optical depth of the dust shell at 0.55~$\mu$m ($\tau$), the dust temperature at the inner radius of the shell ($T_d$), the thickness of the shell $Y = R_{\rm out, d}/R_{\rm in, d}$ where $R_{\rm in, d}$ is the inner radius and $R_{\rm out, d}$ is the outer radius of the dust shell, and the grain size ($a$).

For a given set of parameters, \texttt{DUSTY} computes the emergent spectral energy distribution, which we fit to the observed JWST NIRSpec, MIRI LRS and MIRI imaging observations by $\chi^2$ minimization using the affine-invariant Markov Chain Monte Carlo (MCMC) sampler \texttt{emcee} \citep{Foreman-Mackey2013}. Uniform priors are adopted for all parameters. Convergence of the MCMC chains is verified by visual inspection and by requiring the integrated autocorrelation times to be much shorter than the chain lengths. We adopt the median of the parameter posterior distributions as the best-fit value and its 68\% confidence interval as the uncertainty. The resulting best-fit parameters and their uncertainties are shown in Table \ref{tab:params} and the best-fit model is overlaid with the data in Figure \ref{fig:spec}. The corresponding posterior distributions are shown in Appendix \ref{sec:app_fit}. The inner and outer radius of the shell are derived from the \texttt{DUSTY} output parameters of the best-fit model \citep{Ivezic1999} and shown in Figure \ref{fig:spec}.

The very high optical depth of the dust shell ($\tau \gtrsim 50$) implies substantial obscuration of the central source at optical and near-infrared wavelengths, naturally explaining the non-detection of the source in the G140H band. While the inferred optical depth is higher than estimated from the 2022 photometry ($\tau \approx 21$; D26), the corresponding bolometric luminosity is lower -- indicating continued fading of the source. The increasing optical depth results in a larger fraction of the total luminosity to emerge at long wavelengths; by directly integrating the best-fit \texttt{DUSTY} models, we estimate the fraction of the source luminosity emerging at $> 25\,\mu$m increases from $\approx 1$\% for the progenitor star to $\approx 2.5$\% and $\approx 13$\% for the 2022-2023 and 2025 remnant (observed with JWST), respectively. This indicates that only a small fraction of the total remnant luminosity is emerging at wavelengths longer than the JWST/MIRI coverage.

While the observed SED does not show evidence of a long-wavelength rising component that would indicate a cooler dust shell missed in the MIR observations, we can empirically constrain the presence of such a shell using the observed data. The outer dust temperature in our best-fit model is $\approx 200$\,K; therefore we only investigate possible cooler shells with $T_d \lesssim 100$\,K. As the MIR SED and the effects of dust self-absorption (i.e. the silicate feature) is well described by the warm dust modeled here, we can assume that any outer cooler shell will be optically thin to its own emission. In such a case, the outer cooler SED can be described by a modified black-body function \citep{Fox2010, Fox2011, Myers2024} using optical constants for Silicate dust \citep{Draine2001}. Since a cold dust component would only appear at the longest wavelengths, we estimate the maximum luminosity allowed by scaling the component luminosity to match the observed F2550W flux\footnote{This is a very conservative maximum luminosity estimate since the inner hot dust shell also contributes to the luminosity at this wavelength.}. For a $\approx 100$\,K component, we find the maximum allowed luminosity to be $\lesssim 40$\% of the total measured remnant luminosity at MIR wavelengths. While a $\approx 50$\,K component could still hide $\approx 100$\% of the remnant luminosity as the MIR SED, its implied radius ($\gtrsim 8000$\,au) would be unphysically large (implying ejecta velocities $\gtrsim 4000$\,km\,s$^{-1}$). We therefore rule out a cold dust component that could outshine the observed remnant MIR luminosity.

We follow the relations in D26 to estimate the total dust mass 
\begin{equation}
    M_d = \frac{4 \pi \tau R_{\rm in, d} R_{\rm out, d}}{\kappa_d} \approx 1.4 \times 10^{-4}\,{\rm M}_\odot
\end{equation}
where $\kappa_d \approx 5 \times 10^3$\,cm$^2$\,g$^{-1}$ is the average visual opacity per unit dust mass. The measured dust mass is at the lower end of the distribution observed for core-collapse SNe at $\approx 1000$\,d \citep{Wesson2015, Bevan2019}; however, the measurement is very similar to the measured dust masses in the low-energy mass ejections observed for stellar mergers at similar phases \citep{Karambelkar2025}. As the ejected material is expected to be roughly of solar composition, such condensation is theoretically expected in low-energy transients due to grain condensation from metals \citep{Bermudez2024, Gonzalez2024}.

\subsection{Molecular gas modeling}

\begin{figure*}
    \centering
    \includegraphics[width=0.99\linewidth]{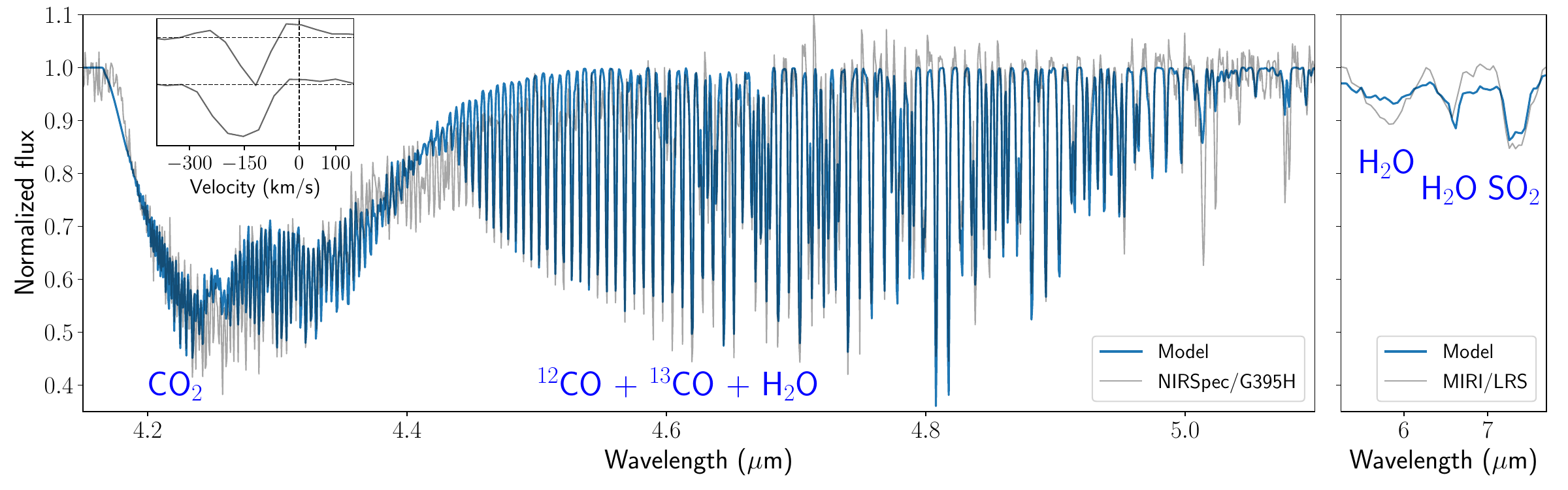}
    \caption{Modeling of the molecular gas features in the NIRSpec and LRS data using a model involving a gas slab in front of the dust photosphere (see text). ({\it Left}) Zoom-in of the best-fit model for the continuum-normalized NIRSpec/G395H data (no significant absorption is detected at shorter wavelengths). The dominant absorption features are highlighted. The inset shows two example velocity profiles (relative to M31 disk velocity $v = -200$\,km\,s$^{-1}$ in the barycentric frame; dashed vertical line) of absorption features showing excess red emission above the continuum (dashed horizontal line), indicative of weak P-Cygni like profiles. ({\it Right}) Corresponding fit in the MIRI/LRS data between $5.2 - 7.7\,\mu$m with the same color scheme.}
    \label{fig:molfit}
\end{figure*}

The MIR spectrum of M31-2014-DS1 exhibits clear evidence for narrow, sharp absorption features between $4-5\,\mu$m in addition to weaker features in the LRS data between $5-8\,\mu$m that is smeared by the low resolution of the data. By comparing to existing literature on dusty Asymptotic Giant Branch (AGB) stars \citep{Cernicharo1998, Cami2002}, we associate the strong features in the NIRSpec band to i) molecular CO$_2$ gas (forming the deep trough at $4.2 - 4.4\,\mu$m), ii) CO gas (including both isotopologues $^{12}$CO and $^{13}$CO) producing the majority of the forest between $4.5$ and $5.0\,\mu$m), iii) water absorption (producing some narrow features in $4.5 -5.0\,\mu$m and unresolved features between $5.0 - 7.0\,\mu$m) that is commonly seen in stellar merger remnants \citep{Lynch2004, Lynch2007, Karambelkar2025} and iv) SO$_2$ absorption that is observed in oxygen-rich environments of AGB stars \citep{Justtanont2004}. The absorption dominated spectrum (without strong emission features) indicates that the absorbing gas is spatially confined very close to the hot (inner) dust continuum source that dominates this wavelength range \citep{Rybicki1979}. To therefore model the gas absorption on top of the inner dust shell, we first derive an accurate continuum by fitting a polynomial function to the overall dust emission between $3-8\,\mu$m.

Figure \ref{fig:molfit} shows the NIRSpec and MIRI LRS molecular absorption between $4-8\,\mu$m after dividing the data by the resulting continuum model. We assume a physical model of a gas slab consisting of CO (both isotopologues of $^{12}$CO and $^{13}$CO), CO$_2$, H$_2$O and SO$_2$. We use the \texttt{slabspec} module of the \texttt{spectools-ir} package \citep{Salyk2022} and the HITRAN \citep{Rothman2010} database\footnote{\url{https://hitran.org/hitemp/}} to compute the wavelength-dependent optical depth (assuming local thermodynamic equilibrium) -- sampled on the same wavelength grid and spectral resolution as the data -- of a given molecule $X$ given its column density ($N_X$), temperature ($T_X$) and velocity broadening ($\delta v_X$). We perform a $\chi^2$ minimization of the resulting transmission model spectrum simultaneously on the NIRSpec and MIRI/LRS data using the \texttt{emcee} code assuming uniform priors. We also include the common systemic velocity of the gas ($v$) as a free parameter (Table \ref{tab:params}). We caution that the low resolution of the MIRI data severely limits estimation of a continuum model due the presence of narrow features; therefore, the MIRI data is not as well fit by the model. However, the dominant molecules in that region (H$_2$O, SO$_2$) do not significantly affect our interpretation.

\begin{table}[!h]
    \centering
    \scriptsize
    \begin{tabular}{lc}
    \hline
    \hline
    Parameter &  Value \\
    \hline
         \texttt{DUSTY} shell model \\
         Total flux ($F$ [erg\,cm$^{-2}$\,s$^{-1}$]) & $(4.09^{+0.05}_{-0.05})\times 10^{-13}$\\
         Composition & Astronomical silicate\\
         Optical depth at $0.55\,\mu$m ($\tau$) & $58^{+6}_{-7}$ \\
         Grain size ($a$ [$\mu$m]) & $0.05^{+0.01}_{-0.01}$ \\
         Inner dust temperature ($T_d$ [K]) & $820^{+60}_{-70}$ \\
         Inner source temperature ($T_*$ [K]) & $5300^{+500}_{-700}$ \\
         Shell thickness ($Y$) & $5.2^{+0.5}_{-0.5}$\\
         \hline
         \texttt{slabspec} molecular gas model\\
Velocity ($v$ [km\,s$^{-1}$]) & $-343.3^{+0.5}_{-0.5}$\\
$^{12}$CO column density ($\log N_{\mathrm{CO}}$ [cm$^{-2}$])
& $21.15^{+0.48}_{-0.45}$\\
$^{12}$CO velocity dispersion ($\delta v_{\mathrm{CO}}$ [km\,s$^{-1}$])
& $5.89^{+0.42}_{-0.33}$\\
$^{12}$CO temperature ($T_{\mathrm{CO}}$ [K])
& $270^{+33}_{-26}$\\
$^{13}$CO column density ($\log N_{\mathrm{CO}}$ [cm$^{-2}$])
& $18.92^{+0.29}_{-0.16}$\\
$^{13}$CO velocity dispersion ($\delta v_{\mathrm{CO}}$ [km\,s$^{-1}$])
& $6.31^{+0.52}_{-0.63}$\\
$^{13}$CO temperature ($T_{\mathrm{CO}}$ [K])
& $294^{+30}_{-38}$\\
H$_2$O column density ($\log N_{\mathrm{H_2O}}$ [cm$^{-2}$])
& $21.34^{+0.28}_{-0.25}$\\
H$_2$O velocity dispersion ($\delta v_{\mathrm{H_2O}}$ [km\,s$^{-1}$])
& $1.21^{+0.07}_{-0.07}$\\
H$_2$O temperature ($T_{\mathrm{H_2O}}$ [K])
& $293^{+32}_{-27}$\\
CO$_2$ column density ($\log N_{\mathrm{CO_2}}$ [cm$^{-2}$])
& $17.86^{+0.01}_{-0.01}$\\
CO$_2$ velocity dispersion ($\delta v_{\mathrm{CO_2}}$ [km\,s$^{-1}$])
& $9.70^{+0.66}_{-0.56}$\\
CO$_2$ temperature ($T_{\mathrm{CO_2}}$ [K])
& $904^{+17}_{-16}$\\
SO$_2$ column density ($\log N_{\mathrm{SO_2}}$ [cm$^{-2}$])
& $17.25^{+0.06}_{-0.06}$\\
SO$_2$ velocity dispersion ($\delta v_{\mathrm{SO_2}}$ [km\,s$^{-1}$])
& $24.8^{+42.6}_{-20.0}$\\
SO$_2$ temperature ($T_{\mathrm{SO_2}}$ [K])
& $271^{+72}_{-59}$\\
         \hline
    \end{tabular}
    \caption{Best-fit parameters and their $68$\% confidence intervals for the \texttt{DUSTY} model to the dust continuum emission and \texttt{slabspec} model for the molecular gas. We only show free parameters directly used in the model, while some derived parameters are mentioned in Figure \ref{fig:spec} and the text.}
    \label{tab:params}
\end{table}

The best-fit model is shown in Figure \ref{fig:molfit} and the molecular parameters are listed in Table \ref{tab:params}. The dominant molecular column density is in the CO gas and water absorption, as commonly seen in the envelopes of highly evolved O-rich AGB stars \citep{Rayner2009} as well as remnants of stellar mergers \citep{Karambelkar2025, Lynch2004}. While the model provides an overall good fit to the bulk of the molecular features, we caution that it has a number of simplifications -- particularly that of a single temperature gas slab, which may not be applicable for a time-variable source with a (likely) density gradient as in M31-2014-DS1. To convert the column density into masses of the gas components, we note that for a gas shell with a $r^{-2}$ density profile, the gas column density is related to the total gas mass by
\begin{equation}
    M_g = 4 \pi R_{\rm in, g} R_{\rm out, g} N_g m_g
\end{equation}
where $M_g$ is the mass of the gaseous species, $R_{\rm in,g}$ and $R_{\rm out, g}$ are the inner and outer radius of the gas shell, $N_g$ is the column density of the component and $m_g$ is mass of the molecule. We estimate the total hydrogen mass in the shell by assuming a relative number abundance between CO and H$_2$ molecules $f_{\rm CO}$. While this number is not directly measured from the data, estimates in the literature for O-rich circumstellar environments of AGB stars \citep{Groenewegen2017, Ramstedt2008, Olofsson2002} range from $f_{\rm CO} = 8 \times 10^{-4}$ to $2 \times 10^{-4}$ . For this work, we nominally adopt a value of $f_{\rm CO} = 3 \times 10^{-4}$ noting that a different assumption would change the respective estimates by order of unity.

To derive constraints on the radial extent of the gas slab ($R_{\rm in,g}$ and $R_{\rm out,g}$), we leverage the nearly pure molecular absorption spectrum in the data. Geometrically, as the size (radius) of the gas slab is increased, it is expected that resulting spectrum would transition from an absorption-dominated spectrum to an emission-dominated spectrum when the geometric emitting area of the gas slab exceeds that of the background sphere. Numerical calculations show that a pure absorption spectrum constrains the size of the emitting gas slab to $R_{\rm out, g} \lesssim 2\times R_{\rm in, d}$ of the underlying continuum emission, and abruptly transitions to an emission-dominated spectrum above this value \citep{Cami2002}. Therefore, we constrain the gas slab to be confined in the inner region of the dusty sphere in radii ranging from $\approx 40-80$\,au. For the best-fit CO column density (Table \ref{tab:params}) and $f_{\rm CO} = 3\times 10^{-4}$, we estimate a molecular hydrogen gas mass of $M_H \approx 0.07^{+0.13}_{-0.04}$\,\Msun.

The high spectral resolution of the NIRSpec data shows that while the broadening velocities for the individual molecules is similar to those in AGB star envelopes \citep{Cami2002}, the source exhibits a clear systemic blueshift motion of $\approx -340\,$km\,s$^{-1}$ relative to the barycentric frame. While this value is close to the known blueshift of M31 \citep{Zhang2024}, M31-2014-DS1 lies $\approx 7$\,kpc away from the center of M31 and is subject to the galaxy rotation curve. To compare the observed velocity to the expected motion of the star, we follow the approach of \citet{Drout2009} and \citet{Massey2009} for supergiants in M31 and use the seminal work on the rotation curve of M31 \citep{Rubin1970}. We find the expected systemic motion of the star to be $\approx -200 \pm 50$\,km\,s$^{-1}$ where we conservatively adopt the observed maximum scatter in the radial velocity motions of red supergiants (RSGs) in M31 \citep{Massey2009} as the systematic uncertainty.

Comparison of the observed velocity to that expected for the source therefore indicates a net blueshift $\approx 100$\,km\,s$^{-1}$, indicative of an expanding layer of gas approaching the observer. Taking the maximum radial extent of the gas layer (out to $\approx 80$\,au) together with the elapsed time since the start of the disappearance ($\approx 3000$\,d; D26) provides a similar estimate\footnote{We do not attempt to derive exact values for this velocity since the true bulk velocity is likely affected by optical depth effects within the expanding slab.} of $\approx 50$\,km\,s$^{-1}$. Evidence for expanding gas is also seen in the red edges of the absorption profiles, where weak emission features are detected in multiple lines (Figure \ref{fig:molfit}), although a detailed modeling including optical depth effects in an expanding atmosphere is beyond the scope of this paper. While evidence for mass ejection and partial fallback presented in D26 was only indirectly inferred from the long-term bolometric luminosity decay (given the lack of an optical outburst), these observations confirm the presence of ejected material. We discuss its implications for the nature of the source in Section \ref{sec:modeling}. 

\section{Modeling}
\label{sec:modeling}

We have thus far presented an interpretation agnostic modeling of {\it JWST} observations of the remnant of M31-2014-DS1. In this section, we aim to interpret the derived parameters (Table \ref{tab:params}) to constrain the nature of the disappearance of the star and test the failed SN hypothesis. Specifically, we aim to explain i) the confirmed bolometric fading of the star to $\approx 7-8$\% of the progenitor luminosity, ii) the extremely dust enshrouded remnant with $\sim 10^{-4}$\,\Msun  of dust located in a shell at $\approx 40-200$\,au, iii) a central heating source for the dust shell with an inferred temperature and blackbody radius of $\approx 5250$\,K and $\approx 100$\,\Rsun, iv) a shell of expanding gas producing the molecular absorption, containing $\sim 0.1$\,\Msun of gas expanding at $\approx 100$\,km\,s$^{-1}$ at a radius of $\approx 40-80$\,au and v) the non-detection of X-ray emission to $L_X \lesssim 1.5 \times 10^{35}$\,erg\,s$^{-1}$. 

\begin{figure*}
    \includegraphics[width=0.49\textwidth]{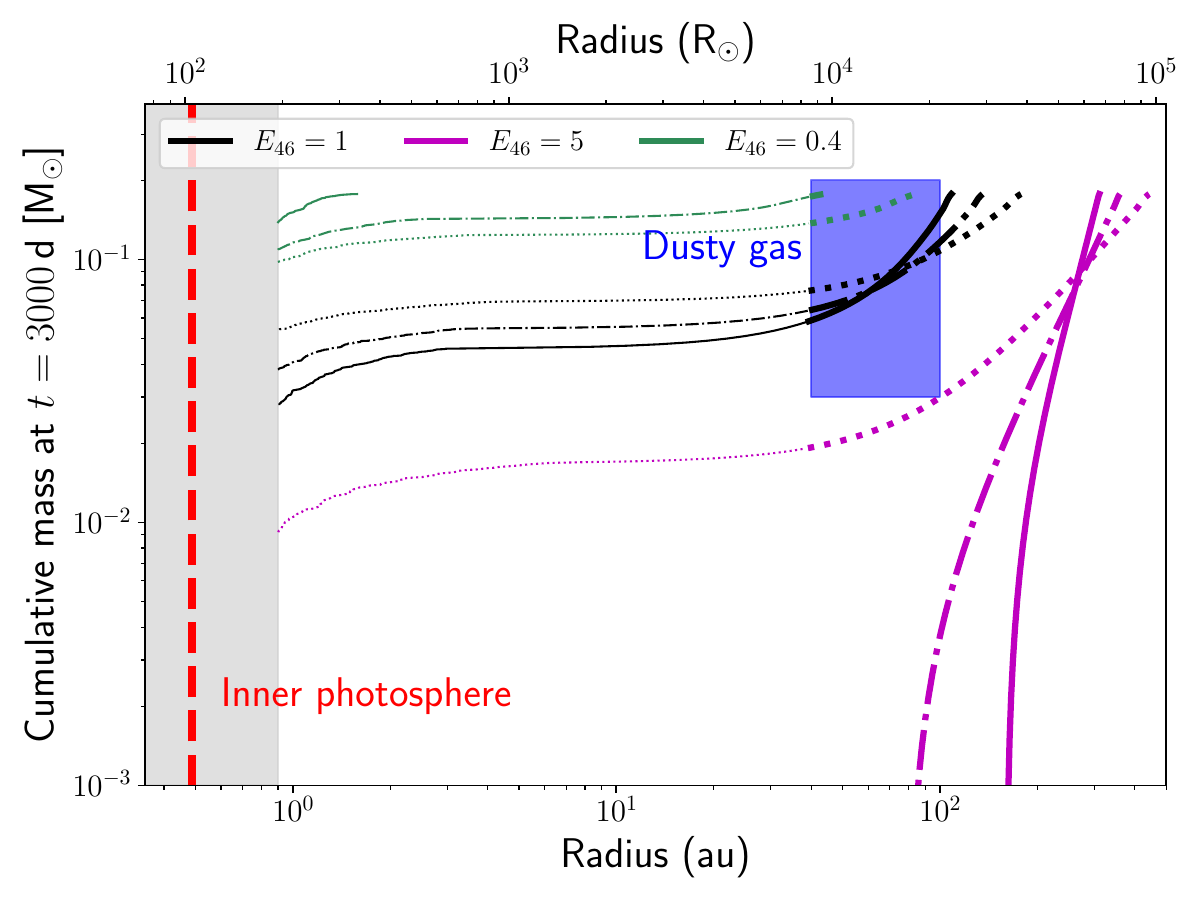}
    \includegraphics[width=0.49\textwidth]{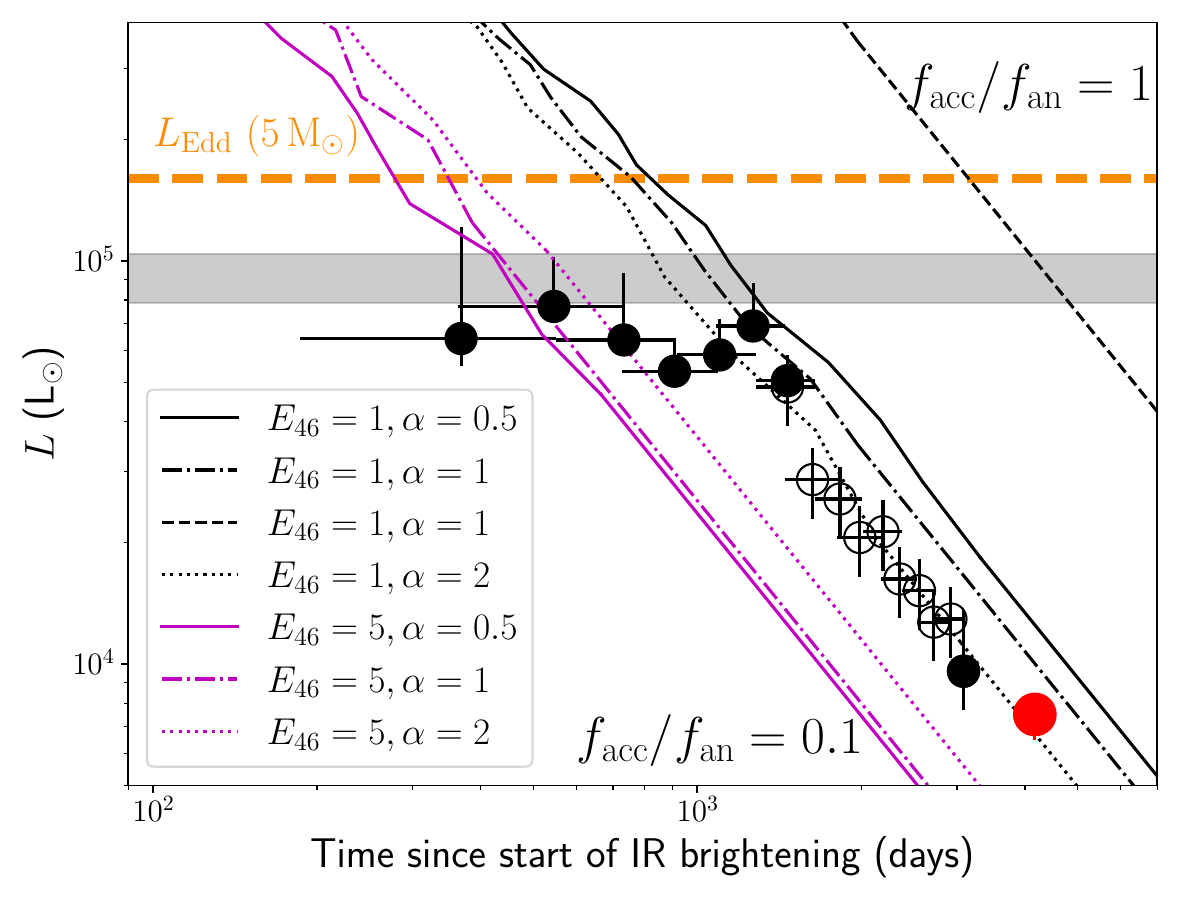}
\caption{(Left) Comparison of the cumulative gas mass profile inferred from JWST observations to model predictions for mass ejection and fallback due to impulsive energy injection. The blue shaded region shows the estimated mass and radius of the gas shell surrounding the remnant and the red vertical dashed line shows the inferred radius of the inner photosphere. The solid, dashed and dotted lines show the predicted cumulative radial mass profile for three different energies ($E_{46} = 0.4, 1, 5$) with line styles indicating the adopted velocity profile as in the right panel. The bold sections indicate radii where the material is unbound (total energy $> 0$), while thin sections indicate bound material. The mass profiles at $\lesssim 1$\,au (shown in the gray region) are not tracked in the simulation since they have fallen back. (Right) Comparison of the long-term bolometric decay of M31-2014-DS1 to our model for late-time mass accretion. The black and hollow circles indicate bolometric luminosity estimates from SED modeling and bolometric corrections to WISE data reported in D26, respectively. The red point denotes the measurement from JWST reported in this paper. The gray shaded region shows the luminosity estimate for the progenitor star and the orange horizontal dashed line shows the Eddington luminosity ($L_{\rm Edd}$) for a $5$\,M$_\odot$ BH. We adopt the same explosion time and radiative efficiency ($5$\%) as in D26. The line styles are as in the left panel (see legend); the case of $E_{46} = 0.4$ is not shown since there is nearly no unbound material. An accretion efficiency of $f_{\rm acc}/f_{\rm an} = 0.1 $ is nominally adopted in the models, but we show an example of $f_{\rm acc}/f_{\rm an} = 1$ for the case of $E_{46} = 0.4, \alpha = 1$ as the black dashed line.}
    \label{fig:model}
\end{figure*}

\subsection{A low kinetic energy mass ejection}

The progenitor of M31-2014-DS1 was classified as a warm YSG based on multi-wavelength SED modeling from $\approx 0.4 - 25\,\mu$m (D26) -- indicating a photospheric radius of $\approx 500$\,\Rsun and a surrounding dust shell with an inner radius of $\approx 110$\,au. In contrast, the remnant is enshrouded in a dust shell formed at much closer radii, confirming new dust formation from material ejected during the disappearance, as suggested by D26. However, unlike D26 who assumed that the majority of the ejected gas had condensed into dust with dust-to-gas mass ratio as in the interstellar medium ($d/g \approx 0.01$), our measurements suggest $d/g \sim 10^{-3}$. This indicates that the ejected material has not completely condensed into dust -- explaining i) the increasing dust optical depth between $\approx 2022 - 2025$ as dust formation continues and ii) the differing temperatures of the gas and dust species (Table \ref{tab:params}) as the material has not reached equilibrium. However, we caution that $d/g$ is subject to uncertainties in $f_{CO}$ and precise dust characteristics (degeneracies in grain size and composition).

We estimate the required kinetic energy ($E$) of the explosion to explain the presence of $\sim 0.1$\,\Msun of molecular gas at $\approx 40-80$\,au\footnote{Given that the molecular gas likely associated with ejected material is confined to $\lesssim 100$\,au, it is possible that the majority of the outer dust is leftover from the progenitor dust shell.}. We first highlight that the velocity of the molecular slab ($\approx 100$\,km\,s$^{-1}$) with the inferred gas mass indicates $E\sim 10^{46}$\,erg as an approximate order of magnitude estimate. To further establish the estimate with quantitative modeling, we take the YSG progenitor model of D26\footnote{The models are available at \url{https://github.com/dekishalay/M31-2014-DS1/} and at Zenodo \citep{zenodo}.} and inject low energy shocks into the stellar envelope.  As in D26, we adopt a simple, physically motivated model in which the ejecta are characterized by a power-law radial velocity profile defined as,
\begin{equation}
v(r) = v_0 (r/R_*)^{\alpha}
\end{equation}
where $r$ is the radius within the stellar envelope, $\alpha$ parameterizes the radial distribution of kinetic energy, and $v_0$ is a normalization that is set by the total input kinetic energy ($E_{46}$, which is $E$ in units of $10^{46}$\,erg). The total mass of the outer H-rich envelope is $\approx 0.3$\,\Msun, which has a gravitational binding energy of $\sim 1.5 \times 10^{46}$~erg.

Appendix \ref{sec:app_ej} describes our numerical method for computing the radial evolution of the envelope mass -- demonstrating the competing effects of mass ejection and fallback at different values of $E_{46}$. We use this method to compute the radial density profile of material as a function of time -- which is sensitive to $E_{46}$ as well as the assumed $\alpha$. Although the exact time of collapse is not well constrained for M31-2014-DS1 due to the lack of a detected optical outburst (D26), the MIR brightening and optical disappearance of the source suggest that the JWST observations were obtained at a phase of $\approx 3000 - 4000$\,d after collapse. We show the calculated cumulative mass profile of the ejected mass and fallback material\footnote{We use $3000$\,d as the estimated phase of the observation, noting that the radial density profile moves laterally in the figure by $\lesssim 30$\% within the range of possible times since the ejected material is traveling freely at this phase.} for different values of $E_{46}$ and $\alpha$ in Figure \ref{fig:model}. A higher kinetic energy unbinds a larger fraction of the envelope and places the ejected gas at a larger radii at a fixed time after explosion. A higher $\alpha$ leads to a larger amount of energy in the high velocity layers such that a larger fraction of the envelope falls back at a fixed energy.

Figure \ref{fig:model} shows that the inferred location of the gas layer is consistent with ejection energies of $\approx 10^{46}$\,erg\footnote{The low energy is consistent with the complete lack of an optical outburst prior to the fading (D26, \citealt{Antoni2025}).}. While an ejection energy that is $\gtrsim 3\times$ lower produces too much fallback (leaving little circumstellar material at $\gtrsim 40$\,au), a $\gtrsim 3\times$ higher value ejects the material at too high velocities that reaches larger radii ($\gtrsim 200$\,au) in this time span (see also Appendix \ref{sec:app_ej}). We highlight that the physical cause of the energy injection is not directly constrained by these observations; in the failed SN scenario, we expect this to be a combination of the neutrino mass-loss driven shock \citep{Nadezhin1980, Lovegrove2013} and energy injection due to the angular momentum barrier encountered by infalling material (the majority of which falls back at early times; see Figure \ref{fig:fallsim}) from the outer envelope \citep{Quataert2019, Antoni2023, Antoni2025}. 

Figure \ref{fig:model} shows that the $\approx 0.2$\,\Msun of gas is unbound for $E_{46} \approx 1$, consistent with the observed radial expansion of $v\approx 100$\,km\,s$^{-1}$ of the molecular gas. The remainder of the envelope material ($\lesssim 0.05$\,\Msun) is bound to the star and will fall back -- producing a long tail of infalling material extending to $\lesssim 100$\,\Rsun. We compare the mass and radius of this material to that required to produce the optically thick blackbody component inferred in the center of the dust shell. For a spherical shell of radius $R_*$, the optical depth to electron scattering is
\begin{equation}
    \tau_{\rm es} \sim \kappa_{\rm es}\frac{M_*}{4 \pi R_*^2}
\end{equation}
where $\tau_{\rm es}$ and $\kappa_{\rm es} \approx 0.34$\,cm$^2$\,g$^{-1}$ are the electron scattering optical depth and opacity, respectively, and $M_*$ is the mass in the shell. Therefore, the amount of mass needed to produce an optically thick ($\tau_{\rm es} \gg 1$) photosphere at the observed radius $\approx 100$\,\Rsun is $M_* \gtrsim 10^{-5}$\,\Msun. Although the radial profile of the fallback material is not well constrained by our simulations (Figure \ref{fig:model}) and may be eventually expelled due to feedback, the total fallback mass at a phase of $\sim 3000$\,d within $\lesssim 300$\,\Rsun ($\sim 10^{-2}$\,\Msun; Figure \ref{fig:model}) is sufficient to produce an optically thick photosphere if the gas is ionized by the central energy source. We turn our attention to this energy source in the next subsection.

\subsection{Accretion-powered heating from inefficient fallback}

While impulsive energy injection into the stellar envelope is agnostic to the physical scenario driving it, the strongest evidence for the failed SN hypothesis for M31-2014-DS1 (as well as NGC\,6946-BH1) lies in its continued bolometric fading to substantially fainter levels than the progenitor star. As argued in D26 and \citet{Kochanek2024}, the bolometric fading of $\approx 10\times$ observed until 2022 suggests a terminal collapse of the stellar core, implying the birth of a BH from the core-collapse. While the analysis of D26 only include MIR data to $\lesssim 5\,\mu$m, the new JWST observations rule out the presence of a cold dust component at $\gtrsim 20\,\mu$m that could hide the stellar luminosity due to dust obscuration. Figure \ref{fig:model} shows the bolometric light curve combining published measurements with the latest JWST observations -- further confirming the continued bolometric fading of the source and the terminal disappearance of the star.

In the failed SN scenario, the late-time luminosity is powered by accretion of marginally bound material that is gradually accreted into the nascent BH. However, the majority of the material ($\gtrsim 99$\%) is expected to be eventually expelled due to the random angular momentum barrier near the BH horizon \citep{Quataert2019,Antoni2022, Antoni2023}. In the case of purely spherical infall, the emergent luminosity would be much lower \citep{Faran2025}. D26 argued that the bolometric fading of both M31-2014-DS1 and NGC\,6946-BH1 requires that the progenitor be depleted of H in its envelope to explain the fading over timescales of $\lesssim 10$\,yr. This is because the sustained ($\gtrsim 10^5$\,d), highly super-Eddington accretion rates expected for RSG progenitors with massive ($\gtrsim 10$\,\Msun) H-envelopes are expected to produce a near-constant luminosity capped near the Eddington luminosity ($\approx 10^{39}$\,erg\,s$^{-1}$ for a $5-10$\,\Msun BH) for centuries. As the mass accretion rate drops below the Eddington rate after this time, the luminosity would fade with time $t$ as $\propto t^{-5/3}$ on similar timescales (over many centuries for RSG progenitors; \citealt{Fernandez2018, Coughlin2018, Faran2025}). The depleted ($\lesssim 1$\,\Msun) H-envelope was independently confirmed by the secure YSG progenitor classification for M31-2014-DS1 and independently supported by data for NGC\,6946-BH1 \citep{Humphreys2019}.

Adopting the constraints on the ejection energy ($\approx 10^{46}$\,erg) from the gas kinematics, we find that the fallback model presented in D26 leads to much brighter emission and slower fading at a phase of $\sim +10$\,yr than observed (Figure \ref{fig:model}; again decaying with time $t$ as $t^{-5/3}$ at these late phases). However, the model of D26 assumed i) a constant, nominal radiative efficiency of $\eta = 0.05$ and ii) an estimate of the fraction of material ($f_{\rm an} \approx 0.01$) that is truly accreted based on an approximate analytic prescription based on the relative (random) angular momentum of infalling material to that at the BH horizon. In reality, additional effects such as mechanical and radiative feedback from the BH accretion can further reduce the fraction of material that is accreted \citep{Antoni2023, Antoni2025}. As shown in Figure \ref{fig:model}, the observed bolometric fading is consistent with explosion energies of $\approx 10^{46}$\,erg if only a small fraction $f_{\rm acc} = 0.1 f_{\rm an} \sim 10^{-3}$ of the mass is accreted compared to that indicated by the angular-momentum based analytic prescription. We therefore find that either the combination of the angular momentum barrier and mechanical/radiative feedback on the infalling gas leads to highly inefficient accretion of the fallback material such that only $\sim 0.1$\% of the total envelope mass is accreted in the long term or that the radiative efficiency is $\sim 0.5$\% -- much lower than accreting BH X-ray binary systems \citep{McClintock2006}.

Despite the high bolometric luminosity emerging from the accreting BH ($\approx 3 \times 10^{37}$\,erg\,s$^{-1}$ emerging entirely in the IR band), no X-ray emission (expected to emerge from close to the BH horizon) was detected to $L_X \lesssim 2\times 10^{35}$\,erg\,s$^{-1}$. We attribute this to extremely high inferred column density of H-gas from the ejected material -- which is directly measured as $\log N_H ({\rm cm}^{-2}) \approx 24.7$ and alone sufficient to obscure the X-ray emission by a reprocessing layer (D26; see their Figure S10) -- in addition to the fallback material at $\lesssim {\rm few} \times 100$\,\Rsun. While $N_H$ associated with the ejecta should decline with time as $\propto t^{-2}$, D26 approximated that the fallback component becomes dominant ($\propto t^{-5/3}$) at late-times. Given the low inferred ejection energy, the nominal model for the fallback column density (based on the mass density near the turn-around radius; D26) would indicate an even higher contribution of the fallback material to $N_H$ at this phase ($\log N_H ({\rm cm}^{-2}) > 25.5$).

\section{Summary}
\label{sec:summary}

\begin{figure}[!t]
    \centering
    \includegraphics[width=0.99\linewidth]{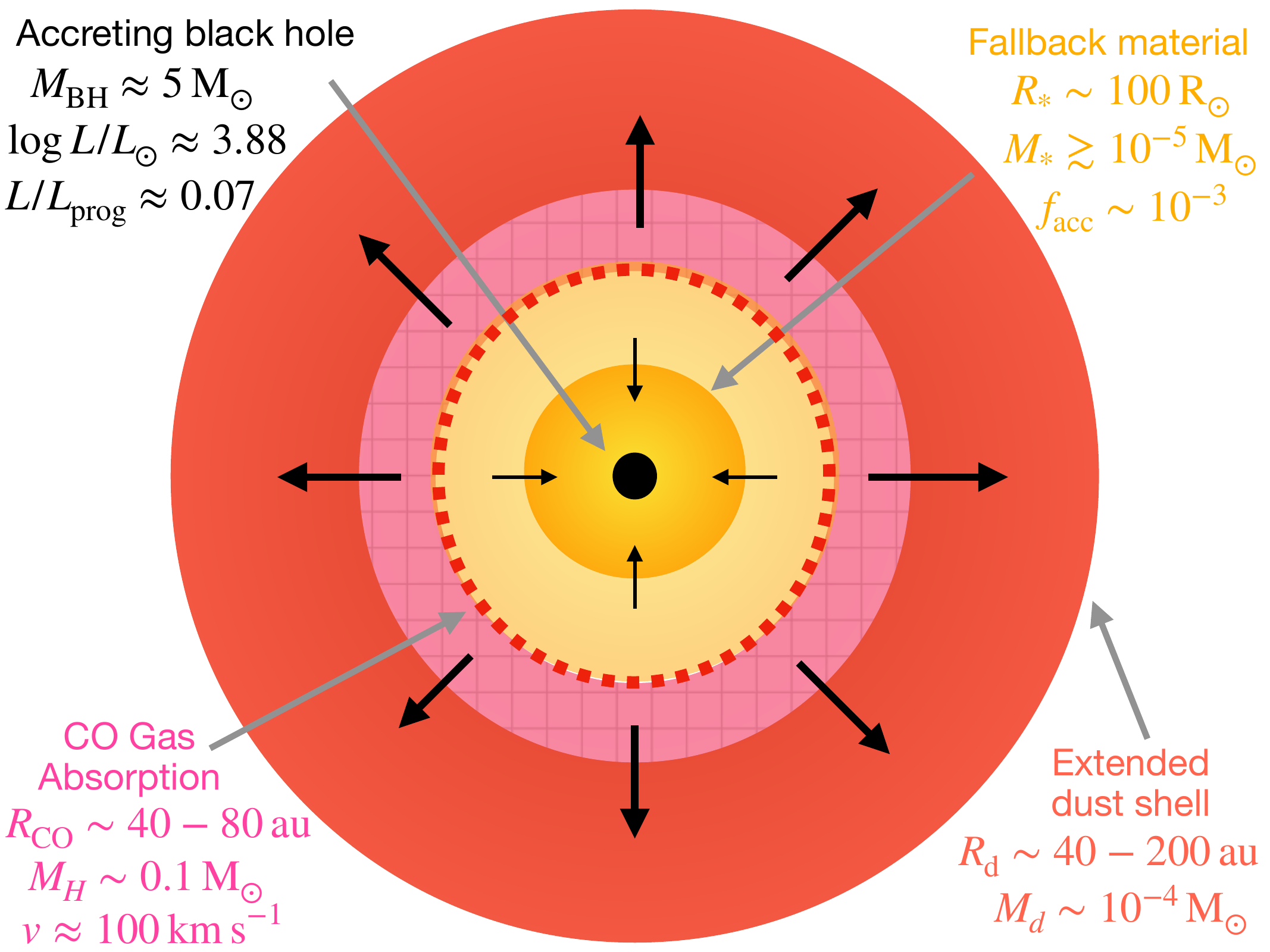}
    \caption{Schematic model of mass ejection and fallback in the remnant of M31-2014-DS1. We show the inferred properties of the gas and dust shell surrounding the remnant, likely consisting of low kinetic energy gas ejected due to neutrino mass loss and inefficient accretion. The black arrows show the inferred direction of motion for the different components. The inferred properties of the inner photosphere and central accreting BH are shown (see text marked by gray arrows). }
    \label{fig:cartoon}
\end{figure}

We have presented multi-wavelength observations from JWST and Chandra of the remnant of the recently discovered failed SN candidate in the Andromeda galaxy, M31-2014-DS1. Our model is summarized in the schematic shown in Figure \ref{fig:cartoon}. By first performing modeling that is agnostic to the physical interpretation followed by association with the failed SN interpretation, we show that 
\begin{itemize}
    \item The remnant is enshrouded in an optically thick dust shell with $\tau \gtrsim 50$, and has become progressively redder since the 2022 observations reported in D26 Comparing to the progenitor properties, we show formation of $\sim 10^{-4}$\,\Msun of new dust at a radius of $\approx 40-200$\,au.
    \item By modeling the narrow absorption features of CO, CO$_2$, H$_2$O and SO$_2$ on the dust continuum, we infer the presence of a close shell of molecular gas located at $\approx 40-80$\,au containing $\sim 0.1$\,\Msun of gas, consistent with the mass of the progenitor H-rich envelope. The molecular features show a net blueshift of $\approx 100$\,km\,s$^{-1}$, allowing us to associate it with expanding ejecta produced by the eruption that enshrouded the remnant.
    \item The source continues to fade dramatically in bolometric light, reaching $\approx 7$\% of the progenitor luminosity in 2024. Together with the $>30$\,yr archival history of constant brightness, we use the continued bolometric fading as evidence for a terminal mass ejection that formed a BH and a slowly fading remnant powered by fallback accretion.
    \item By comparing the inferred location and velocity of the molecular gas to models of mass ejection due to impulse energy injection, we find that the observations can be explained by the injection of $\approx 10^{46}$\,erg shock caused by a combination of the neutrino mass-loss and energy injection from inefficient accretion. When combined with the observed bolometric fading behavior, we infer that the fallback accretion
    is either very inefficient (accreting $\sim 0.1$\% in mass) or that the accretion radiative efficiency is very low ($\sim 0.5$\%).
    \item The accretion luminosity is not detected in X-ray observations to $\sim 100\times$ deeper limits than the IR luminosity. We show that the non-detection is explained by the high column density of ejected material in 2024 -- which is expected to decrease due to expansion of the ejecta, and expected to become dominated by fallback material.
\end{itemize}

While similar quality data is not available for the remnant of NGC\,6946-BH1 (which is much fainter due to its larger distance), the striking similarities between M31-2014-DS1 and NGC\,6946-BH1 suggest that they represent an emerging channel of massive stellar deaths associated with the `quiet' formation of stellar mass BHs as suggested by SN theory \citep{Burrows2025, Janka2025}. The low inferred kinetic energy of the ejection is consistent with the lower range of theoretical predictions for shocks powered by neutrino mass loss \citep{Fernandez2018, daSilva2023, Coughlin2023}, and is somewhat lower than theoretical predictions for explosion energy powered by inneficient accretion \citep{Antoni2023}. Due to its remarkable proximity, M31-2014-DS1 is poised to become the benchmark system for understanding stellar mass BH formation when enabled with long-term follow-up.

Looking ahead, we highlight two promising avenues to confirm and understand this source -- i) under the nominal evolution expected from the ejecta and fallback material, the X-ray source should become optically thin to gas absorption ($N_H \lesssim 24.5$) at $\gtrsim 10^4$\,d after disappearance, suggesting that future X-ray monitoring on the timescale of a decade may reveal the inner accreting BH at the sensitivity level of CXO and the proposed AXIS mission \citep{Koss2025} and ii) as the ejected gas continues to form dust, the future bolometric evolution of the source will be best traced with both continued JWST observations as well as far-IR observations possible with the proposed PRIMA mission \citep{Moullet2025}. More broadly, our results highlight the remarkable capabilities of combining wide-area infrared surveys with the follow-up capabilities of JWST and future IR missions to reveal some of the most poorly understood phases in stellar evolution that are enshrouded in dust.

\newpage
\begin{acknowledgments}
We thank the anonymous referee for a careful review of the manuscript. This work is based on observations made with the NASA/ESA/CSA James Webb Space Telescope. These observations are associated with program \#6809. We thank the STScI director and JWST scheduling team for approving and executing these Director's Discretionary observations. The data were obtained from the Mikulski Archive for Space Telescopes at the Space Telescope Science Institute, which is operated by the Association of Universities for Research in Astronomy, Inc., under NASA contract NAS5-03127 for JWST. Some of the data presented in this article were obtained from the Mikulski Archive for Space Telescopes (MAST) at the Space Telescope Science Institute. The specific observations analyzed can be accessed via \dataset[doi:10.17909/c3sv-qh71]{https://doi.org/10.17909/c3sv-qh71}. This research employs a list of Chandra datasets, obtained by the Chandra X-ray Observatory, contained in~\dataset[doi:10.25574/cdc.538]{https://doi.org/10.25574/cdc.538}.

We thank the CXO director for approving our Director's Discretionary observations and the CXO team for scheduling these observations. This work made use of data obtained from the Chandra Data Archive and software provided by the Chandra X-ray Center (CXC).

We thank A. Boogert for helpful discussions on the gas phase absorption modeling. We thank C. Salyk for helpful discussions on the use of \texttt{spectools-ir}.
\end{acknowledgments}





%
\facilities{JWST (NIRSpec and MIRI), NEOWISE, CXO (ACIS)}

\software{  
          DUSTY \citep{Ivezic1997}, 
          spectools-ir \citep{Salyk2022}, emcee \citep{Foreman-Mackey2013}
          }

\appendix

\section{Parameter posterior distributions from \texttt{DUSTY}}
\label{sec:app_fit}

\begin{figure*}
    \centering
    \includegraphics[width=0.99\linewidth]{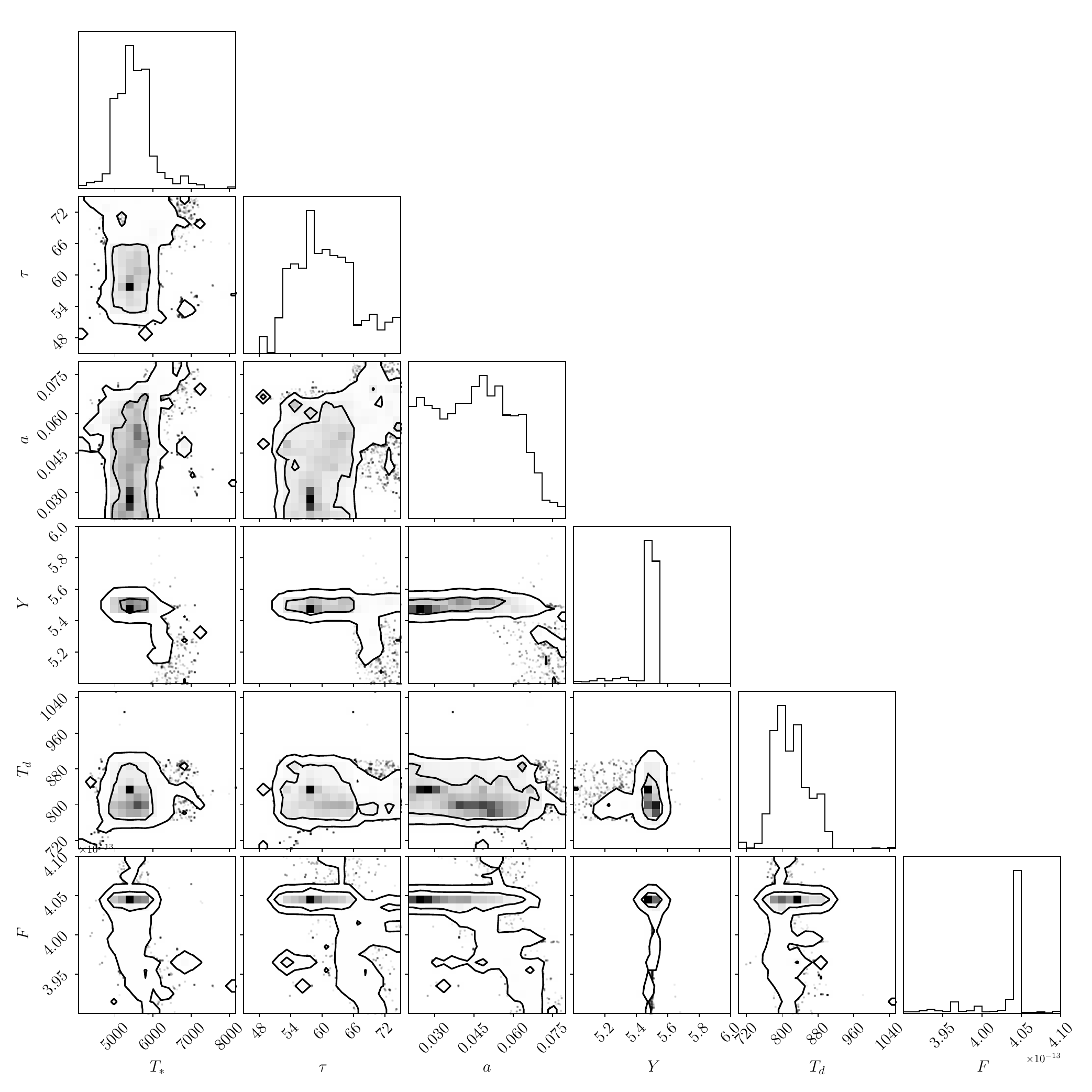}
    \caption{Corner plot showing the posterior distributions of the fit parameters from Table \ref{tab:params}. The lines show the 68\% and 95\% confidence contours.}
    \label{fig:post_dusty}
\end{figure*}

We fit the MIR SED of M31-2014-DS1 using the \texttt{DUSTY} code, as described in Section \ref{sec:analysis}. Here, we show the posterior distributions of the resulting fit in Figure \ref{fig:post_dusty}. The parameters do not show significant degeneracies in the distribution. While some parameters (e.g. the total flux $F$) are very well constrained, the grain size $a$ is not; however, it does not affect our analysis.

\section{Ejection and fallback of material}
\label{sec:app_ej}
\begin{figure*}[!ht]
    \centering
    \includegraphics[width=0.49\linewidth]{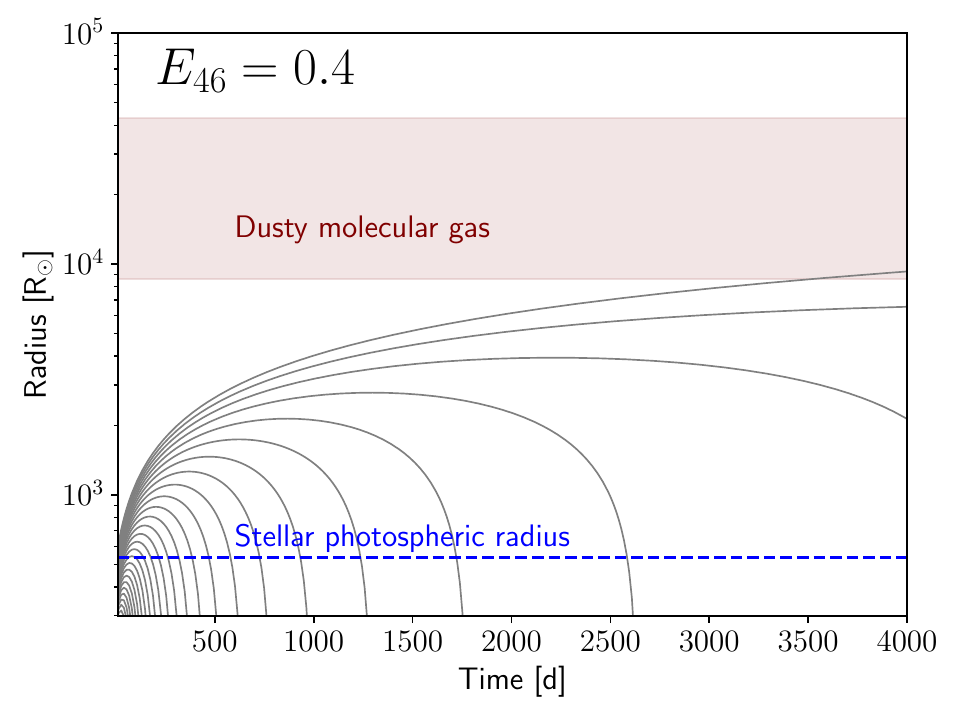}
    \includegraphics[width=0.49\linewidth]{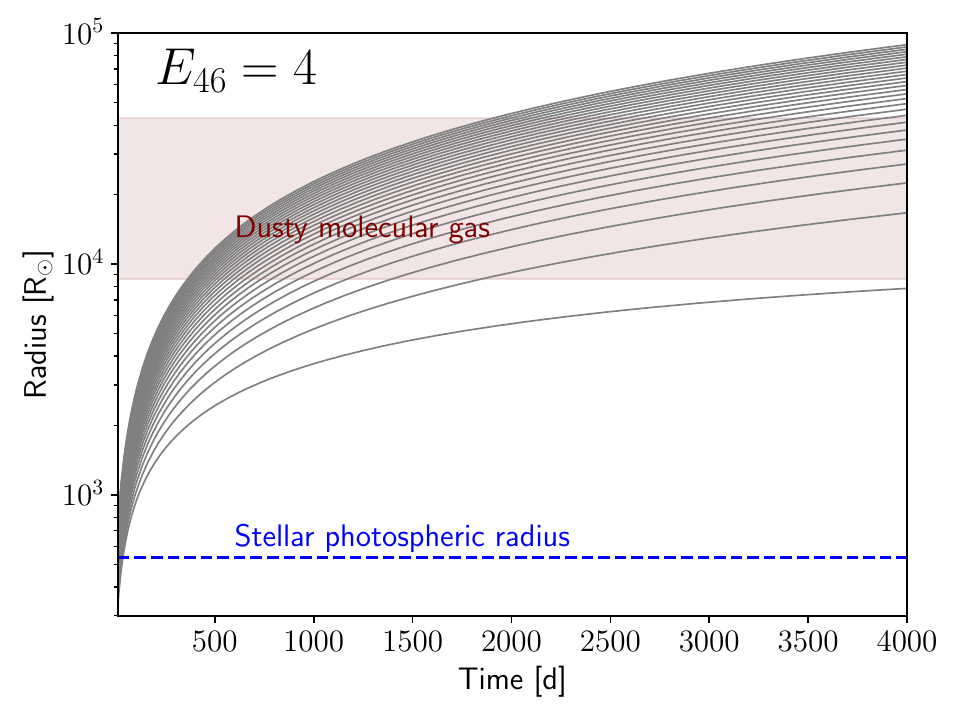}
    \caption{Comparison of trajectories for individual mass shells within the star for different input ejection energies (indicated in the plots) and $\alpha = 1$ --- $E_{46} = 0.4$ (left) and $E_{46} = 4$ (right). The gray lines show radial trajectories as a function of time for material (originally) at different radii within the progenitor envelope. The blue dashed line shows the estimated photospheric radius of the progenitor star and the maroon shaded region shows the estimated location of the dusty molecular gas.}
    \label{fig:fallsim}
\end{figure*}

To estimate the radial density profile of ejected material as a function of time, we construct a one-dimensional, Lagrangian shell model to follow the ballistic evolution of an extended stellar envelope following an impulsive energy injection. The progenitor density and enclosed mass profiles are discretized into a set of concentric radial shells spanning $0.1–1.0\times$ the initial stellar radius. Each shell is assigned a mass from the local density and a radial velocity drawn from the power-law velocity profile. The shells are then evolved independently under the gravitational potential of the enclosed mass, neglecting pressure forces and shell–shell interactions, appropriate for the ballistic flow phase. 

The time evolution of each shell is integrated at high temporal resolution and combined to reconstruct an Eulerian radial profile at a fixed epoch. We divide the envelope into 1000 shells for the simulations; for shells that ultimately fall back, we follow their evolution until their radii become smaller than $\approx 100$\,\Rsun, making the radial profiles reliable only at larger radii (Figure \ref{fig:model}). Figure \ref{fig:fallsim} shows the calculated trajectories of a simulation with 10 shells for two representative cases where the injected energy is smaller and larger than the envelope binding energy ($\approx 7 \times 10^{45}$\,erg). While the majority of the material falls back at late times in the former case, most of the material is ejected to large radii in the latter. From this snapshot we compute shell radii, velocities, densities, and cumulative mass distributions.


\bibliography{sample7}{}
\bibliographystyle{aasjournalv7}



\end{document}